\documentclass[prl,showpacs,twocolumn]{revtex4}
\usepackage{mathrsfs}
\usepackage{stmaryrd}
\usepackage{times}
\usepackage{amsmath}
\usepackage{amsfonts}
\usepackage{amssymb}
\usepackage{graphicx}
\usepackage{bm}
\begin{document}

\title{Majorana Fermion Induced Non-local Current Correlations in Spin-orbit Coupled Superconducting Wires}
\author{Jie Liu$^{1}$} \thanks{Correspondence address to: jliuphy@hku.hk}
\author{Fuchun Zhang$^{1}$}
\author{K. T. Law$^{2}$}\thanks{Correspondence address to: phlaw@ust.hk}

\affiliation{$^{1}$ Department of Physics, and Center of Theoretical and Computational Physics, The University of Hong Kong, Hong Kong, China}
\affiliation{$^{2}$Department of Physics, Hong Kong University of Science and Technology,Clear Water Bay, Hong Kong, China}

\begin{abstract}
Recent observation of zero bias conductance peaks in semiconductor wire/superconductor heterostructures has generated great interest, and there is a hot debate on whether the observation is associated with Majorana fermions (MFs).  Here we study the local and crossed Andreev reflections of two normal leads attached to the two ends of a superconductor-semiconductor wire. We show that, the MFs induced crossed Andreev reflections have significant effects on the shot noise of the device and strongly enhance the current-current correlations between the two normal leads. The measurements of shot noise and current-current correlations can be used to identify MFs.
\end{abstract}


\maketitle

{\bf \emph{Introduction}} --- The search for Majorana fermions (MFs) in condensed matter systems has been an important topic in recent years as MFs are non-Abelian particles [\onlinecite{ivanov, alicea}] and have potential applications in quantum computations [\onlinecite{kitaev,nayak}]. Recent proposals suggest that MFs can appear as zero energy end states in superconducting wires constructed by inducing superconductivity on semiconductor wires with Rashba spin-orbit coupling through proximity effect [\onlinecite{ sau, fujimoto, sato, alicea2, lut, oreg, potter}]. Remarkably, several experimental groups [\onlinecite{kou, deng, das1,marcus}]  recently reported the observation of zero bias conductance peaks (ZBCPs) in Andreev reflection experiments by coupling a normal lead to the end of the aforementioned semiconductor/superconductor heterostructure. These ZBCPs were possibly due to the MF induced Andreev reflections [\onlinecite{law, wimmer}]. However, the origin of these ZBCPs remains a subject of debate [\onlinecite{jie, altland, piku, kells,tewari,pientka, loss, aguado}]. 

\begin{figure}
\centering
\includegraphics[width=3.25in]{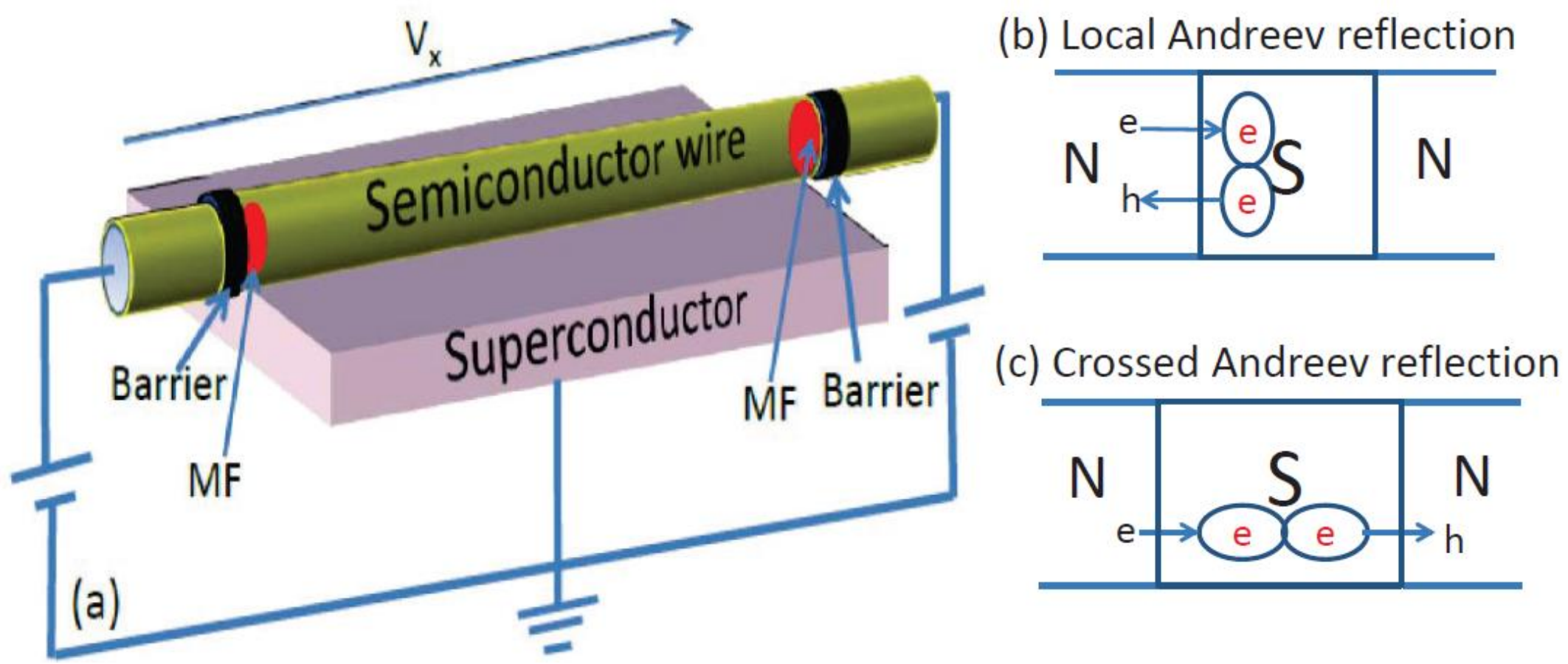}
\caption{ (a) A schematic setup of experiment, two normal leads are coupled to the two ends of a superconductor which supports MFs. (b) A local Andreev reflection process. An electron from one lead is reflected as a hole in the same lead.  (c) A crossed Andreev reflection process. An electron from one lead is reflected as a hole in another lead and a Cooper pair is injected into the superconductor.}\label{f1}
\end{figure}

In this work, instead of studying MF induced local Andreev reflections [\onlinecite{jie, altland, piku, kells,tewari,pientka, loss, aguado}], we explore the non-local properties of MFs. We show that shot noise, which can be used to probe the fractional charges [\onlinecite{sam, pic, rez,dolev}] and their anyonic statistics [\onlinecite{feldman,law2}] in fractional quantum Hall states, can also be used to probe MFs [\onlinecite{law,nilsson}] in superconducting wires. To be specific, an experimental setup depicted in Fig.1a is studied, in which two normal leads are attached to the two ends of a semiconductor wire with Rashba spin-orbit coupling, proximity induced $s$-wave superconductivity and a magnetic field parallel to the wire. In the topological regime, MFs emerge as end states of the superconducting wire.

In the following sections, we show that when MF end states from the two ends of the wire are strongly coupled, local Andreev reflection processes (as depicted in Fig.1b) can be suppressed and the MF end states induce crossed Andreev reflections (CARs), in which an electron from one lead is reflected as a hole in a different lead (as depicted in Fig.1c). Therefore, each normal lead tunnels one electron with charge $e$ into the superconductor in each tunnelling event instead of $2e$ as in local Andreev reflection processes. As a result, the Fano factor of a normal lead, which is the ratio of the shot noise to the average current of the lead, is reduced from $2e$ to $e$ in the CAR regime. Moreover, the current of the two spatially separated normal leads are perfectly correlated in the CAR regime since the two electrons from the two leads have to form a Cooper pair in order to tunnel into the fully gapped superconductor. 

It has been pointed out that several effects such as disorder induced Andreev bounded states [\onlinecite{jie}], Kondo effect [\onlinecite{franceschi, marcus}], weak anti-localization [\onlinecite{piku}] and reflectionless tunnelling [\onlinecite{wees, marmorkos}], may cause zero bias conductance in tunnelling experiments. However, all these effects are essentially the enhancement of local Andreev reflections due to various mechanisms at the interface between a normal lead and a superconductor. These effects cannot cause perfect current-current correlations between two spatially separated leads. Therefore, the measurement of Fano factors and current-current correlations can be used to identify MFs.

Moreover, an experimental setup similar to Fig.1a has been fabricated recently [\onlinecite{das1, das2,marcus}] and the shot noise of the set up in the topologically trivial regime is measured [\onlinecite{das2}]. Therefore, the measurements of the shot noise in the topologically non-trivial regime is experimentally feasible.

{\bf \emph{Model and Formalism}} --- To model the quasi-one dimensional s-wave superconductor with Rashba spin-orbit coupling as shown in Fig.1a,  we use the following tight-binding model [\onlinecite{potter, jie}] with $N_x$, $N_y$ and $N_z$ sites in the $x$, $y$ and $z$ directions respectively:
\begin{eqnarray}\label{model}
 H_{q1D}& =& \sum\nolimits_{\mathbf{R},\mathbf{d},\alpha } { - t(\psi _{\mathbf{R} + \mathbf{d},\alpha }^\dag  \psi _{R,\alpha }  + h.c.) - \mu \psi _{\mathbf{R},\alpha }^\dag  \psi _{\mathbf{R},\alpha } } \nonumber \\ 
&+& \sum\nolimits_{\mathbf{R},\mathbf{d},\alpha ,\beta } { - i{U _R} \psi _{\mathbf{R} + \mathbf{d},\alpha }^\dag  \hat z \cdot (\vec{\sigma}  \times \mathbf{d})_{\alpha \beta }   \psi _{\mathbf{R},\beta } } \nonumber  \\ 
 &+& \sum\nolimits_{\mathbf{R},\alpha ,\beta } { \psi _{\mathbf{R}, \alpha }^\dag [(V_x \sigma_x)_{\alpha \beta} +V_{\text{imp}}(\mathbf{R})\delta_{\alpha \beta}]\psi _{\mathbf{R},\beta } } \nonumber \\
& +& \sum\nolimits_{\mathbf{R},\alpha} \Delta \psi _{\mathbf{R},\alpha }^{\dagger} \psi _{\mathbf{R},-\alpha }^{\dagger}+h.c.
 \end{eqnarray}
Here, $\mathbf{R}$ denotes the lattice sites, $\mathbf{d}$ denotes the three unit vectors $\mathbf{d_{x}}$, $\mathbf{d_y}$ and $\mathbf{d_z}$ which connect the nearest neighbor sites in the $x$, $y$ and $z$ directions respectively. $\alpha, \beta$ are the spin indexes. $t$ is the hopping amplitude, $\mu$ is the chemical potential, $U_{R}$ is the Rashba coupling strength, $V_{x}$ is the Zeeman energy caused by a magnetic field along the wire in the x-direction. $\Delta$ is the superconducting pairing amplitude and $V_{\text{imp}}(\mathbf{R})$ is  the on-site random impurity which is  Gaussian distributed with variance $ \overline{V_{\text{imp}}(\mathbf{R})V_{\text{imp}}(\mathbf{R}')}	=\omega^2 \delta_{\mathbf{R},\mathbf{R}'}$ . In this work, we set $V_x=2\Delta$ such that the superconducting wire can support MF end states by tuning the chemical potential. 

The parameters in the tight-binding model are chosen to match the corresponding values in a recent experiment as done in Ref.[\onlinecite{jie}]. Here, $\Delta=250 \mu \text{e}V$, $t=25\Delta$ and $U_{R}=2.5\Delta$. The dimensions of the wire are $N_x a \approx 1\mu m$, $ N_y a \approx 100nm$ and $N_{z}a \approx 60nm$. The length of the wire is about twice the superconducting coherence length $\xi_{0} \approx t a/\Delta $ and about half the length of the experimental value in Ref.[\onlinecite{kou}]. Due to the short length of the wire, as shown in Fig.2a, the energy of the in-gap states versus the chemical potential exhibits oscillatory behaviour in the topologically non-trivial regime as the two MF end states can couple to each other and the coupling strength is an oscillating function of the chemical potential [\onlinecite{loss, sarma}].

To study the current-current correlation mediated by the MF end states, two semi-infinite normal leads are attached to the two ends of the superconductor as shown in Fig.1a. The two normal leads are described by Eq.\ref{model} by setting $\Delta$ to zero. The tunnelling barriers are simulated by the reduced hopping amplitudes $t_{L}= t_{R} = 0.4t$ between the leads and the superconductor where $t_{L}$ ($t_{R}$) denotes the hopping amplitude from the left (right) lead to the superconductor. However, $t_{L/R}$ controls the width of the conductance peak and it is chosen that the width of the conductance peak is about $0.05\Delta$ as shown in the insert of Fig.2b which is larger than thermal broadening width $k_B T \approx 0.02\Delta$ such that the conductance peak cannot be washed out by finite temperature effects.

We use the recursive Green's function method to calculate the scattering matrix of the model \cite{lee} where the scattering matrix is related to the Green's functions of the superconducting wire by
\begin{equation}
S_{ij}^{\alpha \beta}=-\delta_{i,j} \delta_{\alpha,\beta }+i[ \Gamma_{i}^{\alpha} ]^{1/2}*G^r*[ \Gamma_{j}^{\beta}]^{1/2}.
\end{equation}
Here, $S_{ij}^{\alpha,\beta}$ is an element of the scattering matrix which denotes the scattering amplitude of a $\beta$ particle from lead $j$ to an $\alpha$ particle in lead $i$. $i,j=1$ or $2$. 1 and 2 denote the left and the right  lead respectively.  $\alpha, \beta, \in \{e,h\}$ denotes the electron ($e$) or hole ($h$) channels. $G^{r}$ is the retarded Green's function of the superconducting wire. $\Gamma_{i}^{\alpha}=i[(\Sigma_{i}^{\alpha})^r-(\Sigma_{i}^{\alpha})^a]$, where $(\Sigma_{i}^{\alpha})^{r(a)}$ is the $\alpha$ particle retarded (advanced) self-energy of lead $i$.  

With the scattering matrix, the average current $\bar{I}_{i}$ of lead $i$, the differential shot noise $P_{ij}$ and shot noise $C_{ij}$ can be calculated as:\cite{sun, datta}
\begin{equation} {\label{transport}}
\begin{array}{l}
\bar{I}_i=\frac{{\text{e}}}{{h}}\int_0^{\text{e}V} {\sum\limits_{j,\alpha } {\text{Tr}[I{ - \text{sgn}(}\alpha ){S}_{ij}^{{e}\alpha } (E)^{\dag } S_{ij}^{e\alpha } {(E)]}} } dE, \\ 
P_{ij}(E)=\frac{{2{\text{e}}^{2} }}{h}\sum\limits_{\alpha ,k\beta  \ne l\beta '} {\text{sgn}(}\alpha {)\text{Tr}[S}_{il}^{e\beta {'}} {}^{\dag } S_{ik}^{e\beta } {}S_{jk}^{\alpha \beta } {}^{\dag } S_{jl}^{\alpha \beta {'}}(E) {]}, \\
C_{{ij}}=\int_0^{eV} {P_{ij} (E)} {dE},
\end{array}
\end{equation}
where $\text{sgn}(\alpha)=1$ if $\alpha =e$ and $\text{sgn}(\alpha)=-1$ if $\alpha= h$. In this work, we set the chemical potential of the two normal leads to be the same and the voltage bias between the leads and the superconductor to be $V$. Physically, $C_{ij}=\int_{-\infty}^{+\infty}\overline{\delta I_{i}(0) \delta I_{j}(t)} dt $ measures the current fluctuation of leads $i$ and $j$, where $\delta I_{i}= I_{i}(t) - \bar{I}_{i}$ denotes the deviation of the current at time $t$ with respect to the average current $\bar{I}_{i}$. At low temperatures with $k_B T \ll eV$, the current fluctuation is dominated by the shot noise [\onlinecite{blanter}] and $C_{ij}$ is reduced to the shot noise. On the other hand, $\frac{dC_{ij}}{dV}=eP_{ij}$ is the differential shot noise caused by electrons with incident energy $E$.
 
\begin{figure}
\centering
\includegraphics[width=3.25in]{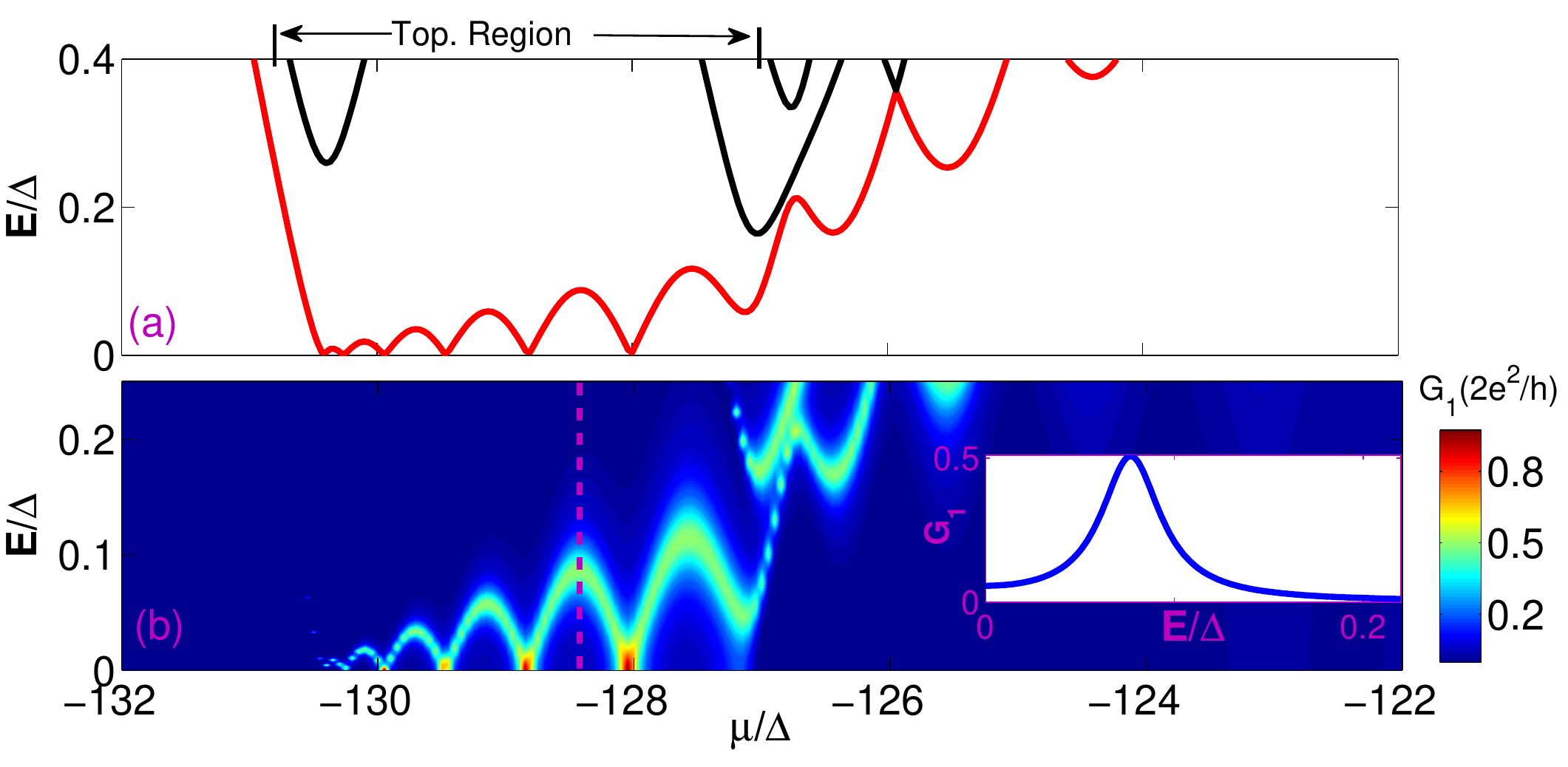}
\caption{ (a)The energy eigenvalues of a short quasi-1D wire versus chemical potential, the lowest energy states are indicated in red. The topological region is marked out. (b) Contour plot of differential conductance $G_1$ of the left lead as a function of chemical potential and electron incident energy $E$. The insert shows $G_{1}$ versus incident energy at a fixed chemical potential denoted by the vertical dashed line in (b). The height of the peak at $E=E_M$ is $ \frac{2e^2}{h}\frac{t_{L}^2}{t_{L}^2 +t_{R}^2}$.}
\end{figure}

{\bf \emph{Current}}--- In this section, we focus on the tunnelling current near the  topological regime where only one transverse subband is occupied and there are MF end states. Due to the oscillatory nature of the MF wavefunctions,  the coupling strength of the MFs oscillate and the resulting coupling energy oscillates as $ E_M \approx \frac{\hbar^2 k_{F}}{m^{*} \xi_{0}} e ^{-2N_x a/\xi_{0}}\cos(k_{F}N_x) $, where $k_{F}$ is the Fermi momentum which is a function of chemical potential and magnetic field and $m^{*}$ is the effective band mass [\onlinecite{sarma}]. The energy spectrum of the superconducting wire is shown in Fig.2a. The topological regime is marked out in Fig.2a [\onlinecite{sup}]. As the chemical potential increases and a second transverse subband is occupied, the superconductor becomes topologically trivial [\onlinecite{potter}].

To study the MF end states, we calculate the differential conductance $d\bar{I}_{1}/dV$ using Eq.\ref{transport}. The contour plot of the differential conductance of the left lead $G_{1}=d\bar{I}_{1}/dV$ as a function of electron incident energy  $E=eV$ and the chemical potential is shown in Fig.2b. As expected, the MFs manifest themselves by inducing conductance peaks. However, in the presence of the second lead and if $E_{M}$ is much larger than the width of the conductance peak, the height of the conductance peak is reduced to $G_1 (E_{M})=d\bar{I}_{1}/dV|_{eV=E_M} \approx \frac{2e^2}{h}\frac{t_{L}^2}{t_{L}^2 +t_{R}^2}$ [\onlinecite{sup}] as shown in the insert of Fig.2b. The numerical calculations in Fig.2b correspond to the case with $t_L = t_R$. As a result, $G_{1}(E_M)=0.5\frac{2e^2}{h}$.

Another interesting point for tunnelling into a superconductor with two strongly coupled MFs is that at low incident energy $E \ll E_{M}$, the differential conductance $G_{1}(E) \propto \frac{2e^2}{h} \frac{t_{L}^2 t_{R}^2}{E_{M}^2}$ depends on the product of $t_L$ and $t_R$ [\onlinecite{sup}]. This means that an electron from a normal lead cannot tunnel into the superconductor unless a second normal lead is present. This is a manifestation of the fact that local Andreev reflection processes are suppressed and the current is purely caused by CAR processes. 

\begin{figure}
\centering
\includegraphics[width=3.25in]{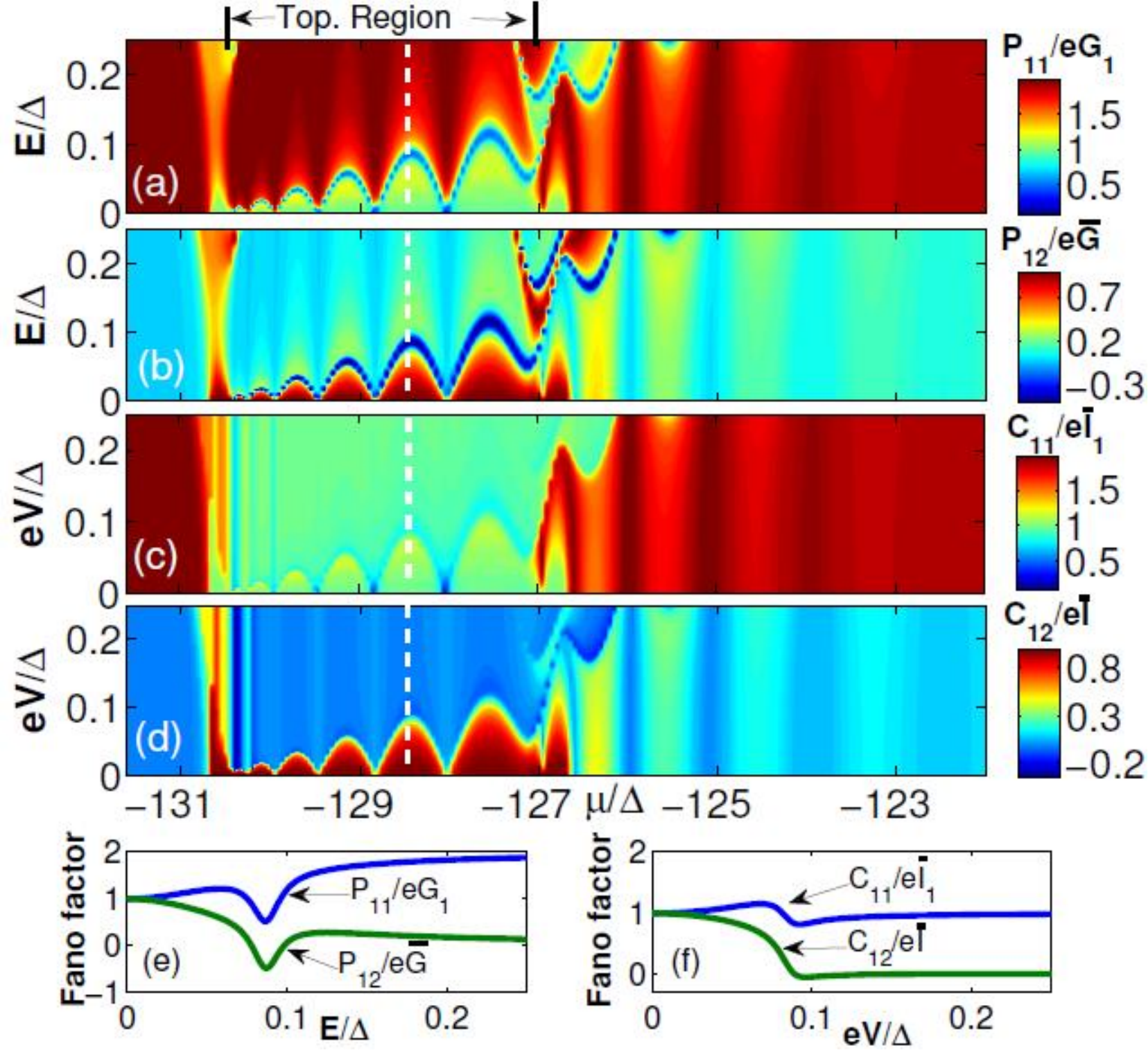}
\caption{ (a) Contour plot of  Fano factor $P_{11}/eG_1$ for electrons with incident energy $E$ at chemical potential $\mu$. (b) Contour plot of $P_{12}/e\bar{G}$. (c) Contour plot of $C_{11}/e\bar{I}_1$ as a function of voltage bias $eV$ and chemical potential. (d) Contour plot of $C_{12}/e\bar{I}$. (e) The $P_{11}/eG_1$ and $P_{12}/e\bar{G}$  as a function of incident energy $E$ at a fixed chemical potential denoted by the dashed lines in (a) and (b), respectively. (f) The $C_{11}/e\bar{I}_1$ and $C_{12}/e\bar{I}$ as a function of voltage bias at fixed chemical potential denoted by the dashed lines in (c) and (d). }
\end{figure}

{\bf \emph{Differential Shot Noise}}--- To probe the MF induced CAR processes, we note that the local and CAR processes can be distinguished experimentally by measuring the Fano factors [\onlinecite{blanter}] of the normal leads. The Fano factor is the ratio of the shot noise to the average current. Physically, the Fano factor measures the electric charge leaving a lead at each tunnelling event given that the tunnelling amplitude is small. 

In this section, we first study the energy dependence of the Fano factors for electrons with incident energy $E$. The contour plot of the Fano factor, $F(E)=P_{11}(E)/G_{1}(E)$, is shown in Fig.3a. Here, $P_{11}(E)$ is the differential shot noise $\frac{1}{e}\frac{dC_{ii}}{dV}$. Here, $P_{11}(E)$ is caused by electrons with incident energy $E$ from the left normal lead, $G_{1}(E)$ is the differential conductance of the left normal lead. Similarly, the crossed current-current correlator $P_{12}(E)$, normalized by the average differential conductance $\bar{G}(E)=\frac{1}{2}(G_1 +G_2)$, is shown in Fig.3b. $P_{12}(E)$ measures the current-current correlations between the left and the right leads. Even though $P_{ij}$ are more difficult to measure experimentally than $C_{ij}$, they give detailed information about different tunnelling processes as a function of $E$ as shown below.

From Fig.3a and Fig.3e, it is evident that the Fano factor $F(E)=P_{11}(E)/G_{1}(E)$ at $E=0$ is the electron charge $e$ in the topological regime. Similarly, it can be shown that the Fano factor for the right lead is $P_{22}(E=0)/G_{2}(E=0)=e$. This indicates that for each tunnelling event, each normal lead contributes one electron in the tunnelling process. Moreover, it is evident from Fig.3b and Fig.3e that $P_{12}(E=0)/\bar{G}=e$. As pointed out in Ref.[\onlinecite{nilsson}], the cross correlator $P_{12}$ is bound by the relation $2|P_{12}| \le P_{11} + P_{22}$ for any stochastic process. At $E=0$, we have  $2|P_{12}(E)|= P_{11}(E) + P_{22}(E)$. This indicates that the two leads are perfectly correlated with each other such that a Cooper pair is injected into the superconductor at each tunnelling event. It is important to note that the almost perfect current-current correlation persists as long as $E \lesssim E_{M}$. This is in sharp contrast to the topologically trivial regime as shown in Fig.3a in which local Andreev reflection processes dominate and the Fano factor for each lead is $2e$. The tunnelling currents of the two leads are only weakly correlated in the absence of MFs as shown in Fig.3b.

It is important to note that the shot noise exhibit strong energy dependence in the topological regime. At $E=E_M$, the Fano factor reaches a minimum value of $\frac{t_{R}^2}{t_{L}^2+t_{R}^2}e$ [\onlinecite{sup}]. For $E \gg E_M$, local Andreev reflection processes dominate and the Fano factor approaches $2e$. On the contrary, the shot noise of $2e$ in the trivial regime is insensitive in the low energy regime.

The cross correlator $P_{12}$ also exhibits strong energy dependence in the topological regime. As shown in Fig.3b and Fig.3e, $P_{12}/e\bar{G} \approx 1$ for $E \lesssim E_{M}$. At $E=E_M$, $P_{12}/e\bar{G} \approx -\frac{2 t_{L}^2 t_{R}^2}{(t_{L}^2 + t_{R}^2)^2}$ [\onlinecite{sup}]. At $E \gg E_M$, local Andreev reflections dominate at each lead and the correlations between the two leads drop to zero eventually. 

{\bf \emph{Shot Noise}}--- In this section, we study the shot noise $C_{ij}$, which is the integration of the differential shot noise over the incident energy as defined in Eq.\ref{transport}. The contour plots of $C_{11}$ and $C_{12}$, normalized by $\bar{I}_1$ and $\bar{I}=\frac{1}{2}(\bar{I}_1 + \bar{I}_2)$ respectively, as a function of chemical potential and voltage bias are shown in Fig.3c and Fig.3d. The Fano factor $C_{ii}/ \bar{I}_{i}$ gives the charge leaving lead $i$ at each tunnelling event. As expected, in the CAR regime with $E \lesssim E_M$,  $C_{ii}/ \bar{I}_{i} \approx e$.  In this case, the two leads are almost perfectly correlated as $C_{12}/\bar{I} \approx e$ as shown in Fig.3f.

Outside of the topological regime, $C_{ii}/ \bar{I}_{i} \approx 2e$ as local Andreev reflection processes dominate. Moreover, the cross correlation between the two leads $C_{12}$ is significant only when the incident energy of electrons satisfies $E \ll E_M$ in the topological regime as shown in Fig.3d. 

\begin{figure}
\centering
\includegraphics[width=3.25in]{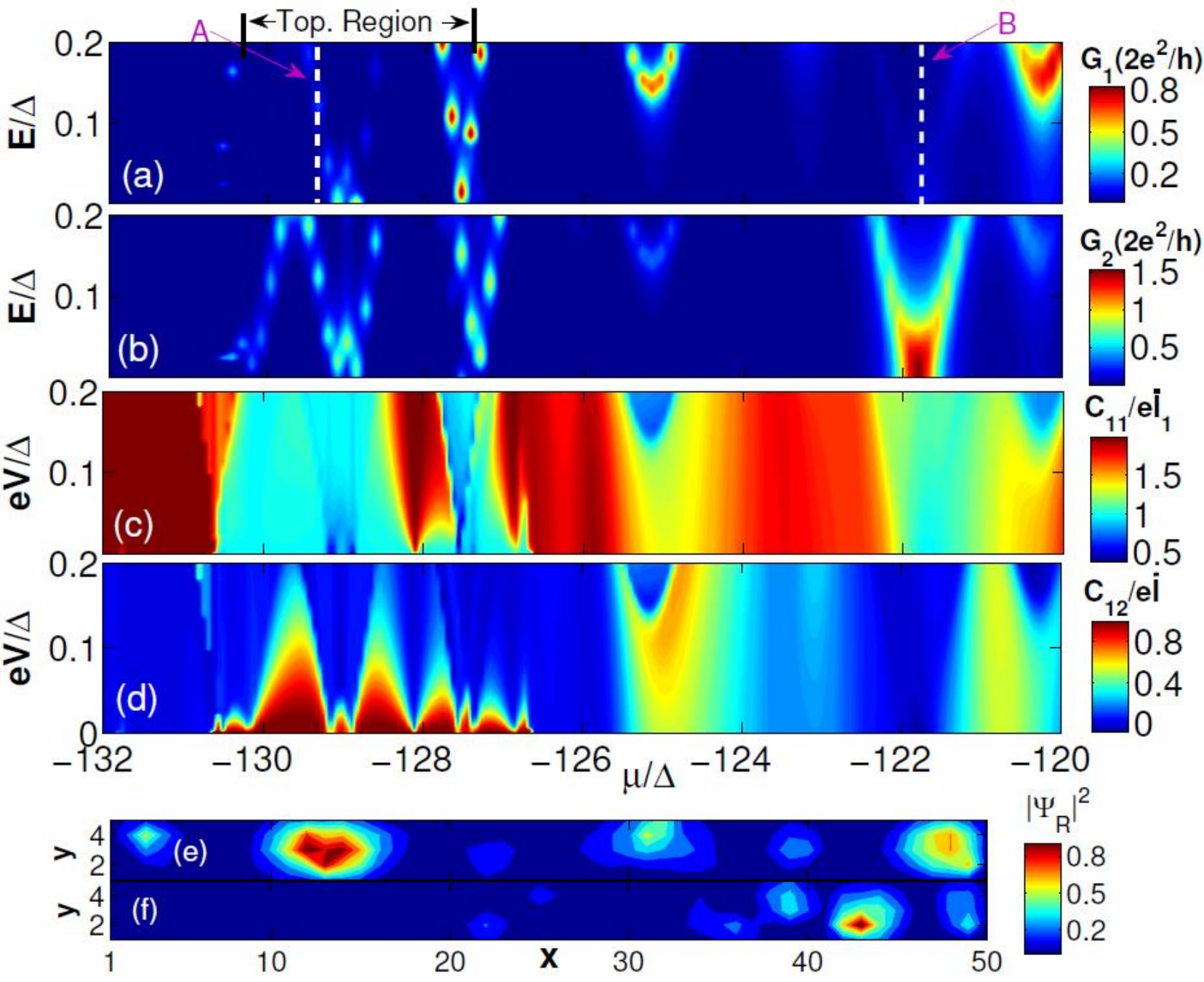}
\caption{Random disorder with $\omega = 16\Delta$ is present for all the figures. (a) and (b): The contour plots of the conductance $G_1$ and $G_2$, respectively. (c) and (d): The Fano factors $P_{11}/e G_1$ and $P_{12}/e\bar{G}$, respectively. (e) The ground state wavefunction $|\Psi _{\mathbf{R}}|^2$ in the topological regime with $\mu = -129.4\Delta$ (indicated by dashed line A in (a)). The dimensions of the wire are $N_x=50a$ and  $N_y= 5a$. (f) The ground state wavefunction in the trivial regime with $\mu = -120.8\Delta$ (indicated by dashed line B in (a)).}
\end{figure}

{\bf \emph{Disorder effect}}--- The observation of the ZBCPs in recent tunnelling experiments is an important step in the search for MFs[\onlinecite{kou,deng,das1,marcus}]. However, as argued in Ref.[\onlinecite{jie}], fermionic end states formed by two MF end states in the topologically trivial regime can also induce ZBCPs in the presence of disorder. Therefore, it is important to distinguish the fermionic end states from the true MF end states. In this section, we show that the shot noise caused by a localized fermionic end state and the shot noise caused by two spatially separated MF end states are different.

On-site random disorder with Gaussian distribution with $\omega=16 \Delta$ is added to the superconducting wire. The contour plots of the differential conductance $G_1$ and $G_2$ for the left and right leads are shown in Fig.4a and Fig.4b respectively. It is important to note that in the topologically trivial regime where two transverse subbands of the wire are occupied, a fermionic end state which has energy close to zero is induced by disorder at $\mu \approx -122\Delta$. The ground state wavefunction at $\mu=-121.8\Delta$ (projected onto the x-y plane) is shown in Fig.4f and it is evident that the ground state is localized at the right end of the wire. As expected, this zero energy fermionic end state induces a strong conductance peak for the right normal lead-superconductor junction. Therefore, it is difficult to distinguish this fermionic state from a true MF end state by measuring the conductance alone.

However, since the fermionic state at $\mu \approx -122\Delta$ is a localized state, the cross current-current correlation $C_{12}/e\bar{I}$ induced by this state is small as shown in Fig.4d. On the contrary, $C_{12}/e\bar{I}$ is close to 1 in the topological regime at $E \ll E_{M}$. Moreover, the Fano factor $C_{11}/e \bar{I}_1$ at $E \approx 0$ is close to 1 only in the topological regime. The ground state wavefunction in the topological regime at $\mu=-129.4\Delta$ is shown in Fig.4e. It is evident that this fermionic end state, which can mediate CARs, is a non-local fermionic state and its wavefunction has significant distribution at both ends of the wire.  Therefore, the experimental signatures  of $C_{12}/e\bar{I} \approx 1$ and $C_{11}/e \bar{I}_1 \approx 1 $ at $E \approx 0$ can be used to distinguish MFs from local fermions.

{\bf \emph{Conclusion} }--- We show that the MF induced CARs change the shot noise and strongly enhance the cross current-current correlations between two leads. The measurements of the Fano factor $e$ of the leads and the strong current-current correlation at small voltage bias can be used to detect MFs. The effects of different magnetic field strength and sample size as well as the transport properties in the multi-subband regime are discussed in the Supplementary Material. It is shown that the MF enhanced CAR effect discussed in this work is very robust and independent of the details of the parameters used.

{\bf \emph{Acknowledgement}}--- We thank H. Barranger, D. Feldman, H. Jiang, P.A. Lee, X.J. Liu, A.C. Potter, Q.F. Sun and especially M. Heiblum for insightful discussions. KTL and JL thank the support of HKRGC through Grant 605512 and HKUST3/CRF09. FCZ thanks the support of RGC HKU707211 and AOE/P-04/08.

\section{Supplementary Material}

{\bf \emph{ Effective Hamiltonian}} --- When a single transverse sub-band of the superconducting wire is occupied, the wire is in the topological regime with two Majorana end states, we expect the transport properties of a N/TS/N junction at $eV \ll \Delta$ can be qualitatively described by the effective Hamiltonian $H_{eff}= H_{L} + H_{M} + H_{T} $, where
\begin{equation} \label{eff}
\begin{array}{l}
H_{N}  = -iv_{f} {\sum\limits_{\alpha \in {L/R} }} \int_{-\infty}^{+\infty}{\psi_{\alpha}^{\dag}(x)\partial_x \psi_{\alpha}(x)} {dx}, \\
H_M  = i E_M \gamma_{L}\gamma_{R} \\
H_{T}  = -i [ \tilde{t}_{L}\gamma_L(\psi_{L}^{\dag}(0)+\psi_L(0))+ \tilde{t}_{R}\gamma_R(\psi_{R}^{\dag}(0)+\psi_R(0) )].
\end{array}
\end{equation}
Here, $H_{N}$ is the Hamiltonian of the left and right normal leads, $\psi_{L/R}$ denotes a fermion operator of the left (right) normal lead. $v_{f}$ is the corresponding Fermi velocity of the leads. $H_{M}$ describes the two coupled Majorana fermions, where $E_{M}$ is the coupling strength between the two MF end states  $\gamma_L$ and $\gamma_R$. The coupling between the leads and the MFs are described by $H_{T}$, where the coupling strengths are denoted by $\tilde{t}_L$ and $\tilde{t}_R$ respectively. 

This model was first introduced in Refs.[\onlinecite{sbolech, snilsson}] and the scattering matrix of the Hamiltonian can be found easily using the equation of motion approach [\onlinecite{slaw}]. We denote the incoming states of the electrons and holes with momentum $k$ in the left and right leads by $\psi_{L/R k}(-)$ and $\psi_{L/R -k}^{\dag}(-)$, respectively. The electron and hole scattering states are denoted by $\psi_{L/R k}(+)$ and $\psi_{L/R -k}^{\dag}(+)$, respectively. The scattering matrix $S$ is defined as :

\begin{equation}
\left( {\begin{array}{*{20}c}
   {\psi _{Lk} (+ )}  \\
   {\psi _{Rk} (+ )}  \\
   {\psi ^\dag  _{L - k} (+ )}  \\
   {\psi ^\dag  _{R - k} (+ )}  \\
\end{array}} \right) = \left( {\begin{array}{*{20}c}
   {S^{ee} } & {S^{eh} }  \\
   {S^{he} } & {S^{hh} }  \\
\end{array}} \right)\left( {\begin{array}{*{20}c}
   {\psi _{Lk} (- )}  \\
   {\psi _{Rk} (- )}  \\
   {\psi ^\dag  _{L - k} (- )}  \\
   {\psi ^\dag  _{R - k} (- )}  \\
\end{array}} \right).
\end{equation}
Using the equation of motion method [\onlinecite{slaw}] and following the notations in [\onlinecite{snilsson}], we have:
\begin{equation}
S(E) \equiv \left( {\begin{array}{*{20}c}
   {S^{ee} } & {S^{eh} }  \\
   {S^{he} } & {S^{hh} }  \\
\end{array}} \right) = \left( {\begin{array}{*{20}c}
   {1 + A} & A  \\
   A & {1 + A}  \\
\end{array}} \right),
\end{equation}
where
\begin{equation}
A = Z^{ - 1} \left( {\begin{array}{*{20}c}
   { - i(E + i\frac{{2\tilde{t}_R^2 }}{v_f })\frac{2\tilde{t}_L^2 }{v_f }} & {\frac{- 2 E_M \tilde{t}_L \tilde{t}_R }{v_f }}  \\
   \frac{2 E_M \tilde{t}_L \tilde{t}_R }{v_f } & - i(E +i\frac{2\tilde{t}_L^2 }{v_f })\frac{2\tilde{t}_R^2 }{v_f } \\
\end{array}} \right).
\end{equation}
and $Z=E_{M}^2-(E +i {2 \tilde{t}_R^2 }/{v_f} )(E + i 2\tilde{t}_L^2/ {v_f })$. From the scattering matrix, the local Andreev reflection amplitude for the, say, left lead is $ - i(E + i\frac{{2\tilde{t}_R^2 }}{v_f })\frac{{2\tilde{t}_L^2 }}{v_f } /Z$. When the two Majorana fermions are not coupled and at zero voltage bias with $E_M=0$ and $E=0$, the local Andreev reflection amplitude is 1. This is called resonant Andreev reflections in Ref.[\onlinecite{slaw}]. However, when the two Majorana fermions are strongly coupled with $E_{M} \gg E \quad \text{and} \quad 4\tilde{t}_{L}^2 \tilde{t}_{R}^2/v_{f}^2$, the local Andreev reflection is strongly suppressed. It is interesting to note that when $|E| = |E_{M}| \gg |\tilde{t}_L \tilde{t}_R/v_f |$, the local Andreev reflection amplitude is $t_{L}^2/(t_L^2 + t_R^2)$ and results in a conductance peak of $\frac{2e^2}{h} \tilde{t}_{L}^2/(\tilde{t}_L^2 + \tilde{t}_R^2)$. All these simple analytic results match the numerical results in the corresponding regime very well, as shown in Fig.2b of the main text. For example, in the main text, the effective coupling between the leads to the superconductor are set to be equal such that $\tilde{t}_L=\tilde{t}_R$ and the effective Hamiltonian predicts a conductance peak of $0.5* 2e^2/h $ at $E=E_M$. This is verified in the insert of Fig.2b.

From the scattering matrix, it is evident that the crossed Andreev reflection amplitude is $\frac{{ 2 E_M \tilde{t}_L \tilde{t}_R }}{v_f }/Z$. Therefore, the crossed Andreev reflection is zero if the two Majorana fermions are not coupled when $E_{M}=0$. As shown above, when  $E_{M} \gg E \quad \text{and} \quad 4\tilde{t}_{L}^2 \tilde{t}_{R}^2/v_{f}^2$, the local Andreev reflection is strongly suppressed to the order of $E \tilde{t}_{L}^2/v_{f} E_{M}^2$. However, in this regime, the crossed Andreev reflection is of order $\frac{{ 2 \tilde{t}_L \tilde{t}_R }}{E_M}$. This results in a conductance of order $\frac{2e^2}{h} \frac{\tilde{t}_{L}^2 \tilde{t}_{R}^2}{E_{M}^2}$ at zero voltage bias $E=0$. Moreover, in this regime, it can be shown using the scattering matrix and Eq.3 of the main text that $P_{11}/\bar{G}_{1}|_{E \ll E_M} \approx e$, as shown by the numerical results. It can also be verified that at $E=0$, $2|P_{12}(E)|= P_{11}(E) + P_{22}(E)$ such that the two normal leads are perfectly correlated to each other. This is also consistent with the numerical simulations as shown in Fig.3 of the main text.

It is interesting to note that the elastic co-tunnelling amplitudes from the left lead to the right lead equal the crossed Andreev reflection amplitude. However, since the left lead and the right lead have equal chemical potential and electrons can also tunnel from the right lead to the left lead, there is no net current from the left lead to the right lead and vice versa. Therefore, elastic co-tunnelling processes do not contribute to the net current of the normal leads.

\begin{figure}
\centering
\includegraphics[width=3.25in]{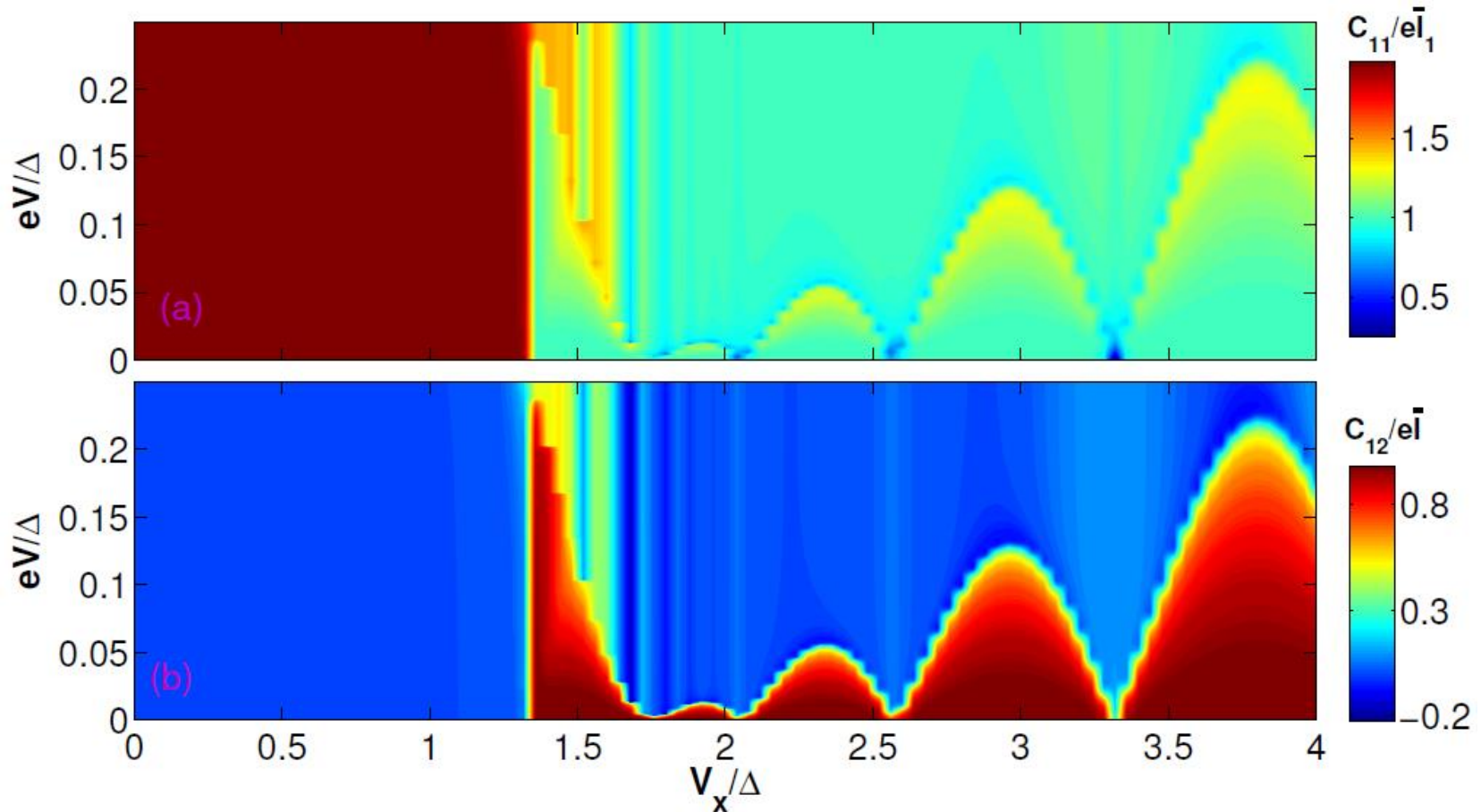}
\caption{(a) Contour plot of $C_{11}/e\bar{I}_{1}$ as a function of voltage bias and magnetic field $V_x$. (b) Contour plot of $C_{12}/e \bar{I}$. } \label{Vx}
\end{figure}

{\bf \emph{Current-current correlations as a function of magnetic field}} --- It is important to note that for the short wire geometry, $E_M$ oscillates as a functions of magnetic field and chemical potential [\onlinecite{sloss, ssarma}]. The energy eigenvalues as a function of the magnetic field strength along the wire are shown in Fig.\ref{Vx}. The chemical potential is $\mu=-130\Delta$. It is evident that when $V_x$ is smaller than the superconducting gap, there are no in-gap states. When $V_x$ is larger than the bulk pairing gap, the superconducting wire is tuned to the topological regime with Majorana end states. Due to the short wire geometry, the coupling energy oscillates as a function of magnetic field. As expected, in when $E_{M} \gg eV$, crossed Andreev reflection processes dominate. 

\begin{figure}
\centering
\includegraphics[width=3.25in]{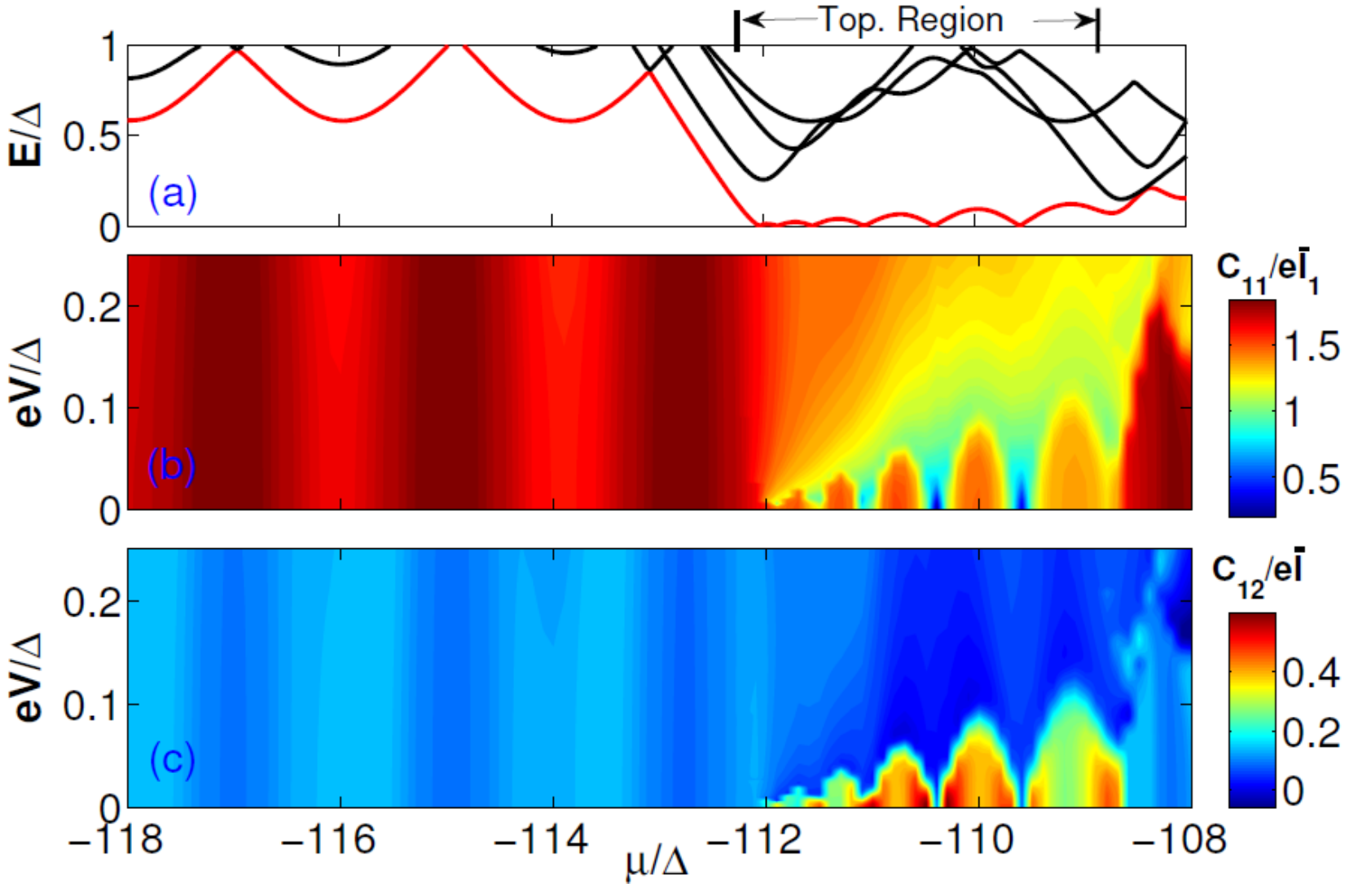}
\caption{(a) The energy eigenvalues as a function of chemical potential. The topological regime with Majorana fermion end states are marked out. (b) Contour plot of $C_{11}/e\bar{I}_{1}$ as a function of voltage bias and chemical potential. (c) Contour plot of $C_{12}/e \bar{I}$. } \label{multi}
\end{figure}

{\bf \emph{Current-current correlations when multiple subbands are occupied}} --- It is important to note that in realistic experiments, it is possible that multi-transverse subbands of the superconducting wire are occupied [\onlinecite{skou,smarcus}]. In the topological regime with an odd number of transverse subbands occupied, and if the number of occupied subbands is larger than one, the appearance of Majorana end states is accompanied by the appearance of finite energy fermionic end states [\onlinecite{spotter}]. Therefore, it is important to show that the measurement of the shot noise and current-current correlations can be used to probe the topological regime even in the presence of other fermionic end states. In this section, we first identify the topological regime by plotting the energy eigenstates of a superconducting wire. When the chemical potential is near the band bottom, only one or two transverse subbands are occupied as shown in the main text. As the chemical potential increases, more transverse subbands of the superconducting wire are occupied. When three subbands are occupied, the wire is again in the topological regime. The energy eigenvalues as a function of chemical potential in the regime where three transverse subbands are occupied are shown in Fig.\ref{multi}a. 

The Fano factor of the left lead $C_{11}/e\bar{I}_{1}$ and the current-current correlations of the two normal leads $C_{12}/e \bar{I}$ are shown in Fig.\ref{multi}a and c respectively. It is evident that in the topological regime with $eV \lesssim E_M $, we have  $C_{11}/e\bar{I}_{1} \approx 1 $ and $C_{12}/e \bar{I} \approx 1$. Therefore, the Fano factors and current-current correlations can still be used to probe the topological regime even in the multi-subband cases. 

{\bf \emph{Identifying the topological regime for short wires}} --- For a wire which is much longer than the superconducting coherence length, the topological regime can be easily identified by plotting the energy eigenvalues versus the chemical potential as done in Fig.2a of the main text. The topological regime is the region where zero energy modes emerge. However, in the short wire geometry discussed in this work, the two Majorana fermions from the two ends may couple and the ground state energy oscillates as a function of chemical potential. As a result, it is difficult to determine the topological regime exactly using the energy eigenvalues. 

\begin{figure}
\centering
\includegraphics[width=3.25in]{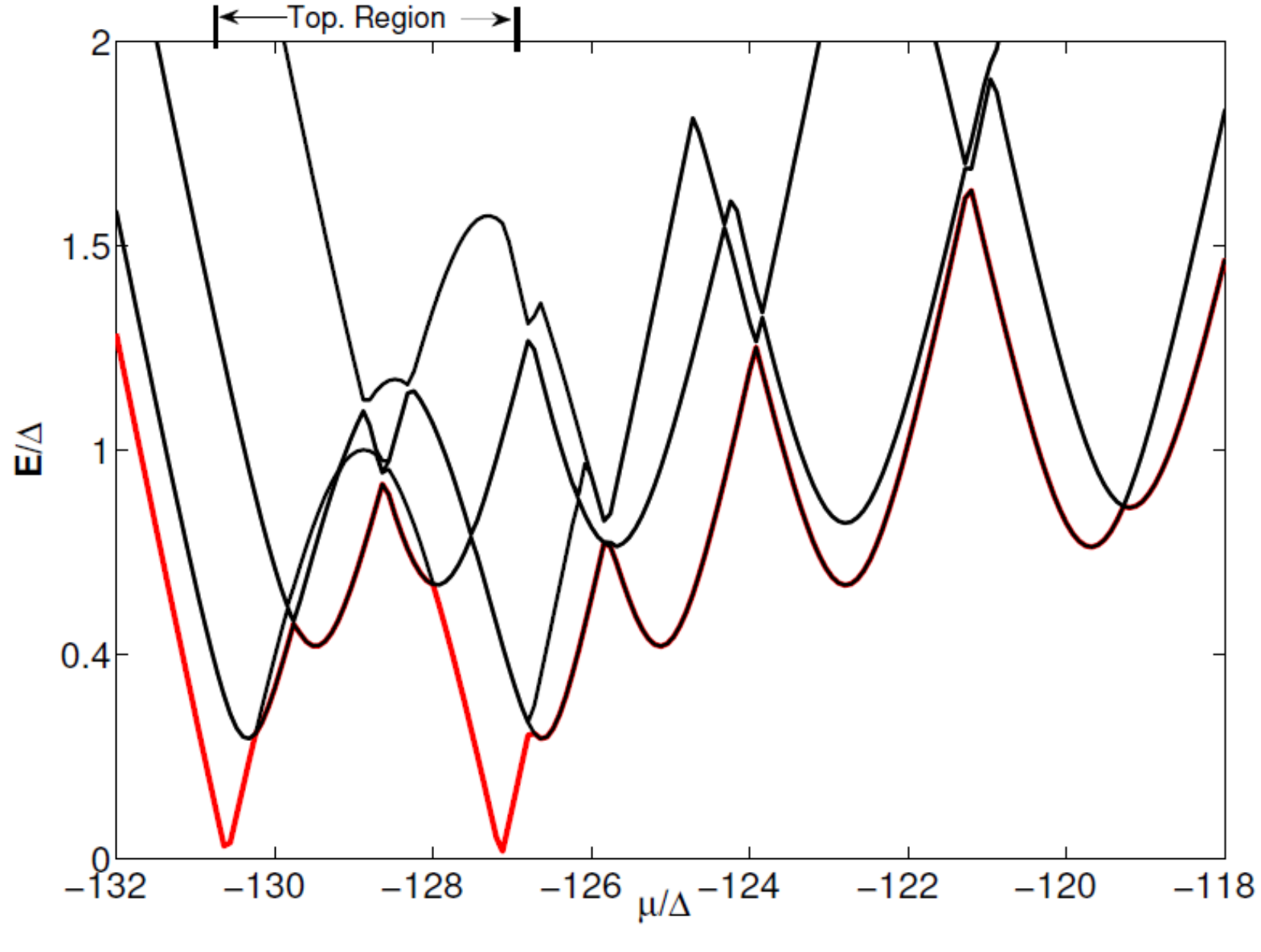}
\caption{ The energy eigenvalues versus the chemical potential for a wire with a ring geometry.  } \label{ring}
\end{figure}

However, due to the fact that the bulk energy gap has to be closed at the topological phase transition point, we can identify the gap closing points as the topological phase transition points. In Fig.\ref{ring}, the energy eigenvalues of a wire with periodic boundary conditions are studied. The ring geometry studied here has the same parameters as the wire studied in Fig.2a of the main text except that periodic boundary conditions in the x-direction are used for the tight-binding model of Eq.1. As the Majorana end states are eliminated in the ring geometry, the gap closing points can be easily found and can be identified as the topological phase transition points. 

\begin{figure}
\centering
\includegraphics[width=3.25in]{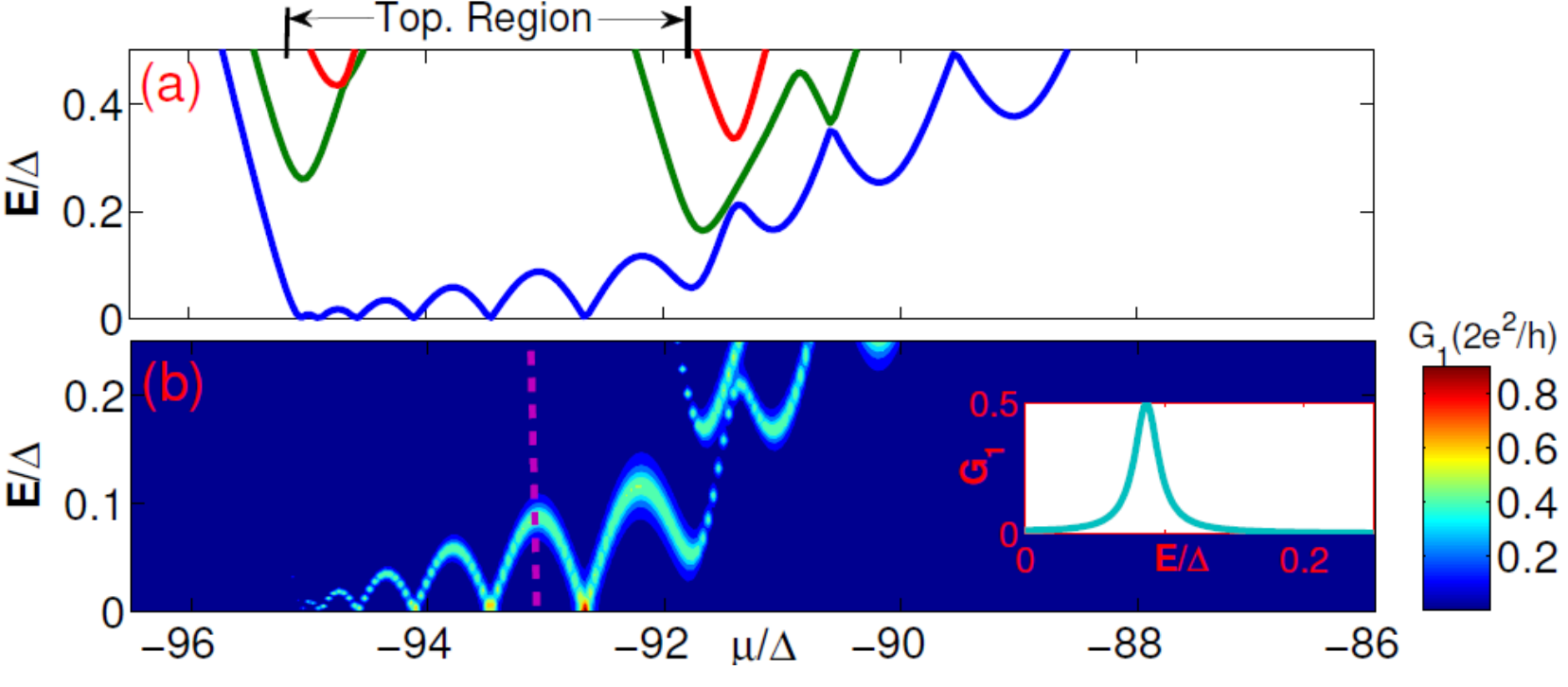}
\caption{ (a)The energy spectrum of a quasi-1D wire with $N_z=1$. The topological region is indicated above. (b) Contour plot of differential conductance $G_1$ of the left lead as a function of chemical potential and electron incident energy $E$. The insert shows $G_{1}$ versus incident energy at a fixed chemical potential denoted by the red dashed line in (b). The height of the peak at $E=E_M$ is $ \frac{2e^2}{h}\frac{t_{L}^2}{t_{L}^2 +t_{R}^2}$. } \label{2Dcurrent}
\end{figure}

\begin{figure}
\centering
\includegraphics[width=3.25in]{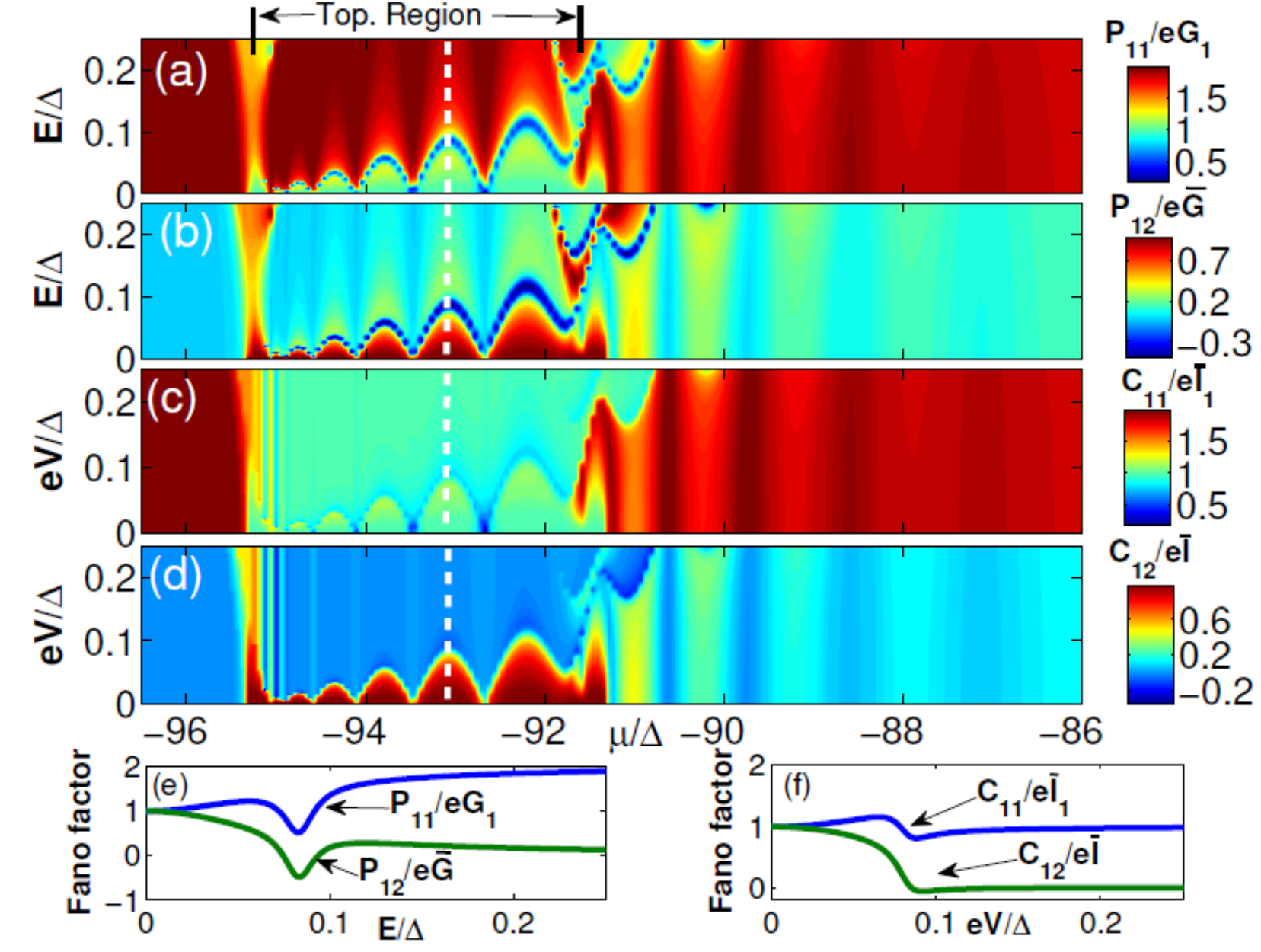}
\caption{ (a) Contour plot of  Fano factor $P_{11}/eG_1$ for electrons with incident energy $E$ at chemical potential $\mu$. (b) Contour plot of $P_{12}/e\bar{G}$. (c) Contour plot of $C_{11}/e\bar{I}_1$ as a function of voltage bias $eV$ and chemical potential. (d) Contour plot of $C_{12}/e\bar{I}$. (e) The $P_{11}/eG_1$ and $P_{12}/e\bar{G}$  as a function of incident energy $E$ at a fixed chemical potential denoted by the dashed lines in (a) and (b). (f) The $C_{11}/e\bar{I}_1$ and $C_{12}/e\bar{I}$ as a function of voltage bias at fixed chemical potential denoted by the dashed lines in (c) and (d). } \label{2Dnoise}
\end{figure}

{\bf \emph{Results using a two dimensional model}} --- It is important to note that the results presented in Fig.2 and Fig.3 of the main text are independent of the details of the model. In this section, a quasi-one dimensional wire with $N_x=50$, $N_y =5$ and $N_z=1$ are used to simulate the superconducting wire. All other tight-binding parameters are chosen to be the same as the ones in Fig.2 and Fig.3 of the main text except for the chemical potential. The hopping strength between the normal lead and the superconductor are chosen as $t_{L/R}=0.3t$ which is different from the $t_{L/R}=0.4t$ in the main text. The energy eigenvalues as a function of the chemical potential and the tunnelling conductance of the left lead are shown in Fig.\ref{2Dcurrent}. The chemical potential is chosen such that only one or two transverse subbands of the wire are occupied. The differential shot noise and the shot noise are shown in Fig.\ref{2Dnoise}. It is evident that in the topological regime where only one transverse sub-band is occupied, the conductance and shot noise exhibit almost identical behaviours as in the case of a 3D wire presented in the main text.

{\bf \emph{The effect of the sample length}} --- It is shown in the main text that the enhanced crossed Andreev reflection regimes occur when the coupling energy between the Majorana fermions is larger than the voltage bias such that $E_{M} \gg eV$. Moreover, the voltage bias has to be larger than the temperature scale $eV \gg k_B T$ such that the shot noise can dominate the thermal noise. Fortunately, these conditions can be easily satisfied in the semi-conductor/superconductor heterostructure. As shown in Fig.2 and Fig.3 of the main text, the maximum coupling energy can be in the order of $0.1\Delta$ which is much larger than $k_B T$. Assuming that $\Delta=0.25meV$ as shown experimentally and $T=20mK$, we have  $k_B T \approx 0.01 \Delta$.

\begin{figure}
\centering
\includegraphics[width=3.25in]{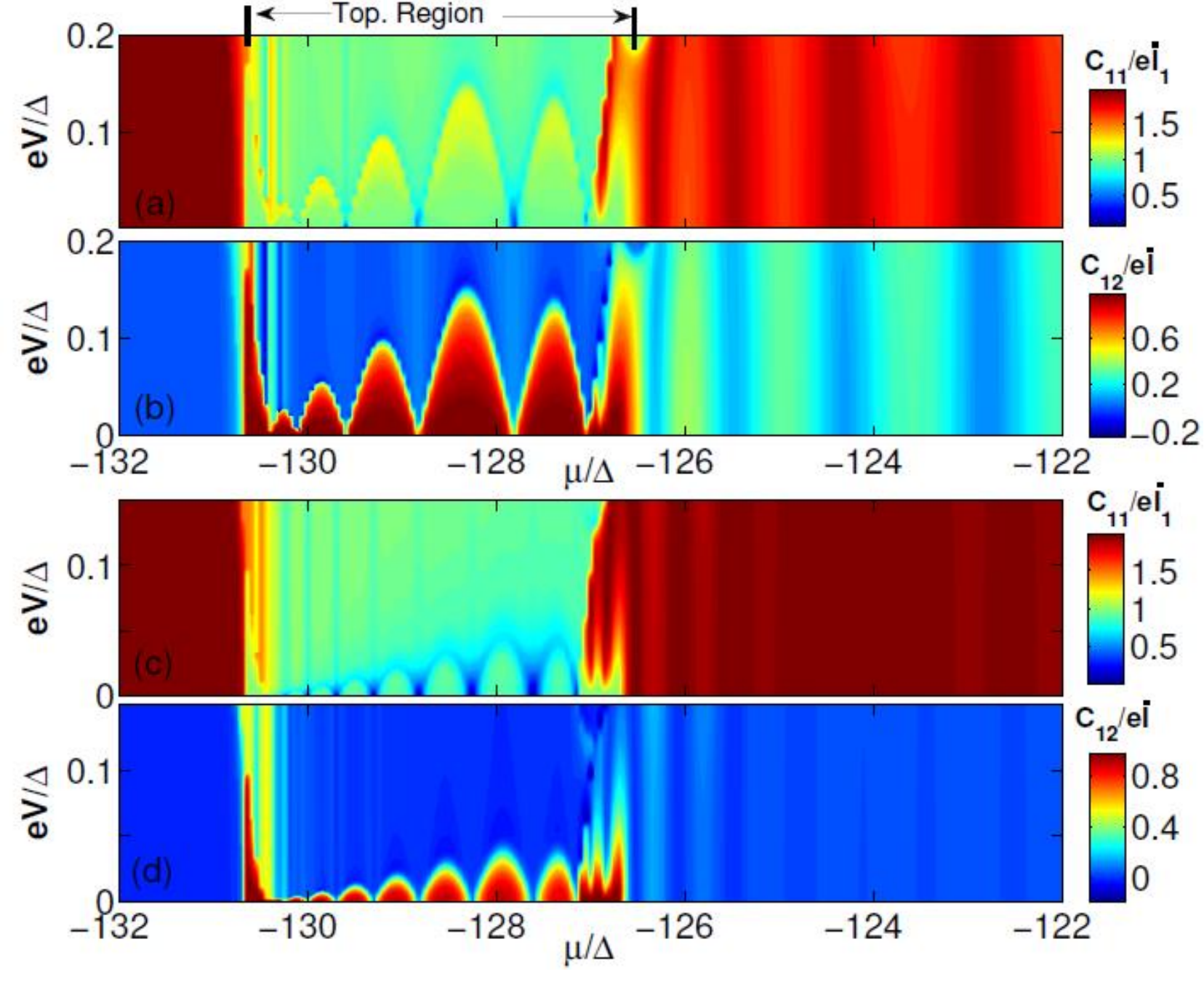}
\caption{ $N_x=40$ for (a) and (b). $N_x=70$ for (c) and (d).  (a) and (c), Contour plots of $C_{11}/e\bar{I}_1$ as a function of voltage bias $eV$ and chemical potential. (b) and (d), Contour plots of $C_{12}/e\bar{I}$. } \label{length}
\end{figure}

Since the coupling strength of the two Majorana end states has the form $ E_M \approx \frac{\hbar^2 k_{F}}{m^{*} \xi_{0}} e ^{-2N_x a/\xi_{0}}\cos(k_{F}N_x) $, the maximum $E_M$ decreases exponentially as a function of the distance between the Majorana end states. The localization length is the superconducting coherence length $\xi_{0} \approx ta/\Delta \approx 25a$. In the main text, a wire length of $N_x a=50a $ is assumed. In this section, results using wires with lengths $N_xa=40a$ and $N_xa=70a$ are presented. As expected, the maximum $E_{M}$ is increased when the wire is shortened. When $N_{x}a=40a$, the maximum $E_{M} \approx 0.15 \Delta$. When $N_{x}a=70a$, the maximum $E_{M}$ is reduced to about $0.05 \Delta$ and the crossed Andreev reflection regime is more difficult to observe. Nevertheless, the length of the wire should not be much shorter than the superconducting coherence length. Otherwise, the current-current correlations between the two leads can be significant even in the absence of Majorana fermions. For example, it is evident from Fig.\ref{length}b and Fig.\ref{length}d that $C_{12}/\bar{I}$ in the topologically trivial regime of the shorter wire with $N_{x}a=40a$ is much larger than $C_{12}/\bar{I}$ in the topologically trivial regime of the wire with $N_xa=70a$.

\end{document}